# Ancient concrete works

Amelia Carolina Sparavigna
Dipartimento di Fisica
Politecnico di Torino

It is commonly believed that the ancient Romans were the first to create and use concrete. This is not true, as we can easily learn from the Latin literature itself. For sure, Romans were able to prepare high-quality hydraulic cements, comparable with the modern Portland cements. In this paper, we will see that the use of concrete is quite older, ranging back to the Homeric times. For instance, it was used for the floors of some courts and galleries of the Mycenaean palace at Tiryns.

**Introduction**
On April 12, 2011, in concurrence with an exhibition in the Oxford's Ashmolean Museum of some ancient treasures of Greece, the BBC announced that Macedonians created cement three centuries before the Romans [1]. Unearthing some tombs of a royal complex belonging to Alexander the Great and his father Philip, at Vergina, archaeologists determined that Macedonians were "not only great warriors but revolutionary builders as well", since they used concrete before the Romans. This information is quite interesting because shows the location of an early use of concrete. Unfortunately, if we want to shed light on the origin of the material, this simple comparison between Macedonian and Roman concretes is a little bit misleading. Moreover, it is raising another question: were Macedonians the first people using concrete? The answer is negative. Concrete was older even than Alexander the Great. Let me show you how we can deduce this fact from some Latin essays and their English translations and from the reports of Schliemann's excavations of Mycenaean archaeological sites.

**The Roman cement**
First, let us remember what concrete is. It is prepared by the hardening of cement and aggregate. The cement is a binder or mortar, which is a substance that sets when mixed with water and then hardens binding together the particles of aggregate. The aggregate could be made of pebbles, ceramic tiles and brick rubbles obtained from the demolition of buildings. It is commonly believed that the ancient Romans were the first to create and use concrete. This is not true, as we can easily learn from the Latin literature itself. For sure, Romans were able to prepare high-quality hydraulic cements, because they used lime and pozzolana, the volcanic dust of Puteoli. The mixture of pozzolana and lime produces a hydraulic binder. The Romans called this binder "caementum". "Opus caementicium" was the Roman concrete. The Roman caementum has much in common with its modern counterpart, the Portland cement: that is, the composition of Roman pozzolana cements is very close to that of modern cement. Probably, it was during the third century BC that hydraulic cement was first prepared by mixing pozzolana with the lime produced by heating limestone [2,3]. The resulting concrete was used for building the harbour at Puteoli (c.199 BC). Let me remember that Puteoli (Pozzuoli), such as Cumae and Naples, before being Roman towns, were Greek colonies of Magna Graecia. May be, the use of concrete could had come from Greece in Italy through these colonies, and, due to the local presence of pozzolana, the hydraulic cement developed and used there, and then, in all the Roman Empire. To the author's knowledge, caementum was the only ancient material similar to the modern Portland cement. It would be interesting to analyse the compositions of Macedonian and Roman concretes to compare their features.

**Formaceos walls**
Early information on the properties of caementum appeared in the books written by Vitruvius and Pliny the Elder [4-6]. Vitruvius devoted a chapter of his book, De Architectura, to pozzolana. Pliny in his Natural History described pozzolana too, but also other concretes had been known in Rome. One was the opus signinum [7], which is obtained from broken pottery, "beaten to powder, and tempered with lime, it becomes more solid and durable than other substances of a similar nature; forming the cement known as the "signine" composition, so extensively employed for even making the pavements of houses". Pliny considered also the "formaceos walls" that can be found in Africa and Spain. He is telling: "they are moulded, rather than built, by enclosing earth within a frame of boards, constructed on either side". For Pliny, "these walls are superior in solidity to any cement. Even at this day, Spain still beholds watch-towers that were erected by Hannibal, and turrets of earth placed on the very summits of her mountains". Even if Pliny is telling that formaceos walls are of "earth", he is probably referring to the tapias [8-10], created with a mixture of lime, sand and gravel, that is, with a mixture for concrete. There is another reference to formaceos walls in a book on the "Rerum Rusticarum" written by Varro, reinforcing the hypothesis of concrete [11,12]. When Varro is discussing the fences used to mark the farm boundaries, he tells that fences can be masonry works too. "There are usually four varieties: those of cut stone, as in the country around Tusculum; those of burned brick, as in Gaul; those of unburned brick as in the Sabine country; those of gravel concrete, as in Spain and about Tarentum". Varro does not describe the components of the mortar.

**The Mycenaean concrete**
Of the English translation of the Varro's book [12], the footnotes are quite remarkable. Who translated and wrote the notes was Fairfax Harrison (1869 – 1938), American lawyer, businessman and writer. In 1913, he became president of the Southern Railway Company. Interesting person: he was an industrialist as well as the writer that translated the agricultural works of ancient Roman writers Cato and Varro [12]. Let me report the complete footnote on Varro's sentence, because it is very important for our discussion. "The kind of fence which Varro here describes as "ex terra et lapillis compositis in formis", is also described by Pliny as formaceos or moulded, and he adds "aevis durant". It would thus clearly appear to have been of gravel concrete, the use of which the manufacturers of cement are now telling us, is the badge of the modern progressive farmer. Cato told how to burn lime on the farm, and these concrete fences were of course formed with lime as the matrix. When only a few years ago, Portland cement was first produced in America at a cost and in a quantity to stimulate the development of concrete construction, engineers began with rough broken stone and sand as the constituents of what they call the aggregate, but some one soon "discovered" that the use of smooth natural gravel made more compact concrete and "gravel concrete" became the last word in engineering practice. But it was older even than Varro. A Chicago business man visiting Mycenae picked up and brought home a bit of rubbish from Schliemann's excavations of the ancient masonry: lying on his office desk it attracted the attention of an engineering friend who exclaimed, "That is one of the best samples of the new gravel concrete I have seen. Did it come out of the Illinois tunnel?" "No," replied the returned traveller, "it came out of the tomb of Agamemnon!"
Really amazing this note written by Fairfax Harrison. It means that Mycenaean people knew and used concrete, before Romans, Carthaginians and Macedonians. Unfortunately, I was not able to find any other detail on the concrete, probably lost forever, of the tomb of Agamemnon. Therefore, I decided to find some hints in the available works written by Heinrich Schliemann, and I found where and for what the Mycenaeans used the concrete.

Besides his famous works at Troy and Mycenae, Schliemann started the excavation of Tiryns, a Mycenaean archaeological site in the Peloponnesus, some kilometres north of Nauplion. Tiryns was a hill fort occupied even before the Bronze Age (see Fig.1-4 for maps and other figures), that reached its height between 1400 and 1200 BC. The fort had a palace and cyclopean walls and tunnels. The detailed reports of the excavations of Tiryns are in a book [13]. There are several sentences in [13] concerning concrete, because people of Tiryns used the concrete for floors. The researchers write that, when excavating and cleaning of the citadel, they "cleared a part of the great gallery to the south-east, of which the upper part forms a pointed arch" (see Fig.3), and "found therein a floor formed of concrete". Moreover, the archaeologists found that the "oldest houses of the primitive settlement" had floors that they found preserved in many places, floors that "were of beaten-down clay and in this quite different from the floor of the palace, built together with the Cyclopean walls on the upper citadel, for this is always of lime concrete".

However, the most interesting description is that of the floors of the men's court of the upper citadel. The book tells: "The floor of the court is composed of a strong concrete of small pebbles and lime, and even now produces a very fine effect. In the middle of the south side, immediately beside the northeast pilaster of the Propylaeum, stands a great altar built of quarried stones. It was apparently dedicated to Zeus, like the altar in the Palace of Odysseus. … Through the Prothyra we enter a great court …. This is the men's court, the centre of the whole palace. The ground-plan is approximately a rectangle, not counting the depth of the colonnades. The whole floor is still covered with thick lime concrete, injured only here and there. In the great shaft at the north-eastern corner of the court, which Dr. Schliemann sunk in 1876, the various strata of this concrete floor are plainly discernible. Lowermost on the remblai, there is a thick stratum (40-70 mm) of stones and lime, a sort of Beton, intended as a secure basis for the actual concrete; then follows a second layer about 15 mm thick, consisting of pebbles and a very solid reddish lime. Uppermost lies a layer of about 18 mm thick, made of lime and small pebbles, and affording a most durable concrete. In some places, especially in the east colonnade, repairs had already been made in ancient times, and with an inferior mortar, consisting almost exclusively of lime. The escape of rainwater is very carefully provided for, for the surface of the concrete is not a horizontal plane, but so levelled that the water runs off to a single point on the south side. There we find a vertical shaft, built of rubble, and covered with a stone flag. Through a hole in this covering stone, the water fell down the shaft, and so reached a walled horizontal canal, which probably led it to some reservoir. No trace of this reservoir, which must have been some metres below the floor of the palace, has yet been found; but its existence may with some confidence be assumed, since the inhabitants of the citadel would hardly have left the water gathered by the roofs and floors to be wasted."

**Conclusion**

As I told at the beginning, cement and concrete are older than Rome and Macedonia. They are so old that are ranging back to the Homeric times. Even the pyramids had their limestone blocks bound together by a mortar of mud and clay [14]. I like to imagine that, at those very ancient times, many cultural patterns ran across the Mediterranean basin, spreading the knowledge of materials and building techniques. As the description of the excavations of Tiryns is showing, the early concrete was used for floors; this means that for the archaeology of concrete it is important to look under our feet.

Fig.1: Maps of Tiryns from [13].

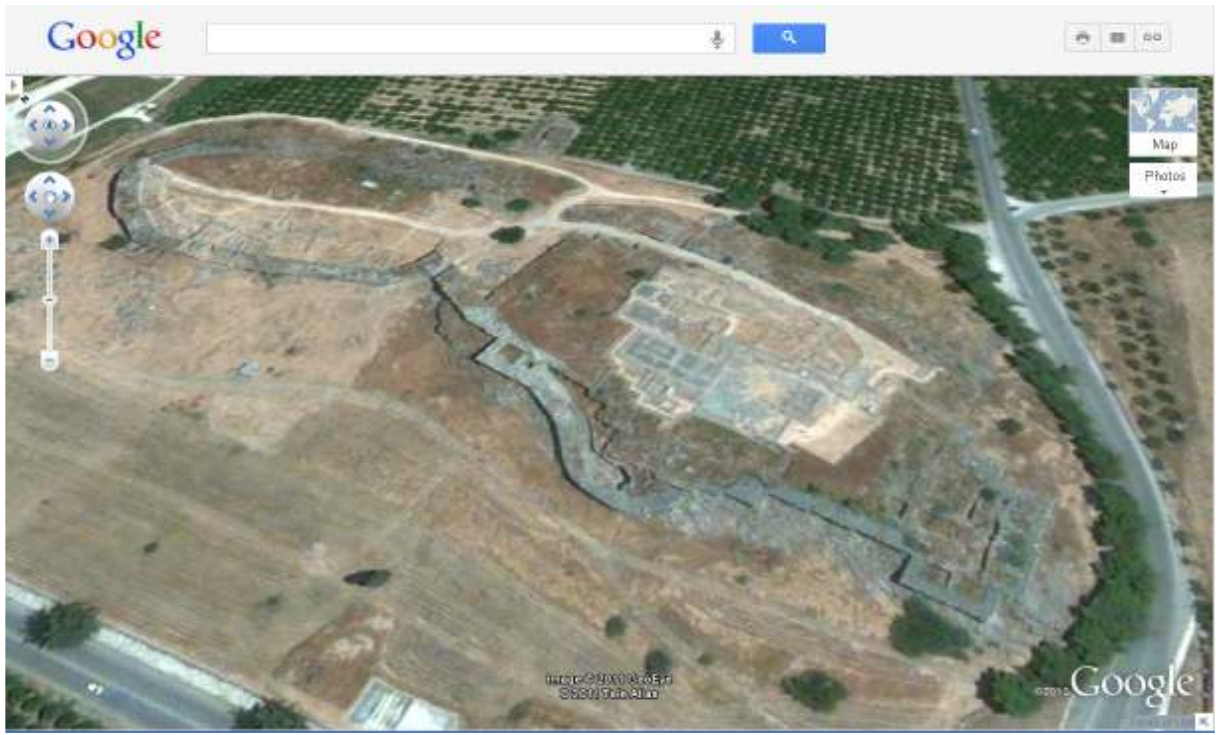

Fig.2: Google Earth view of Tiryns.

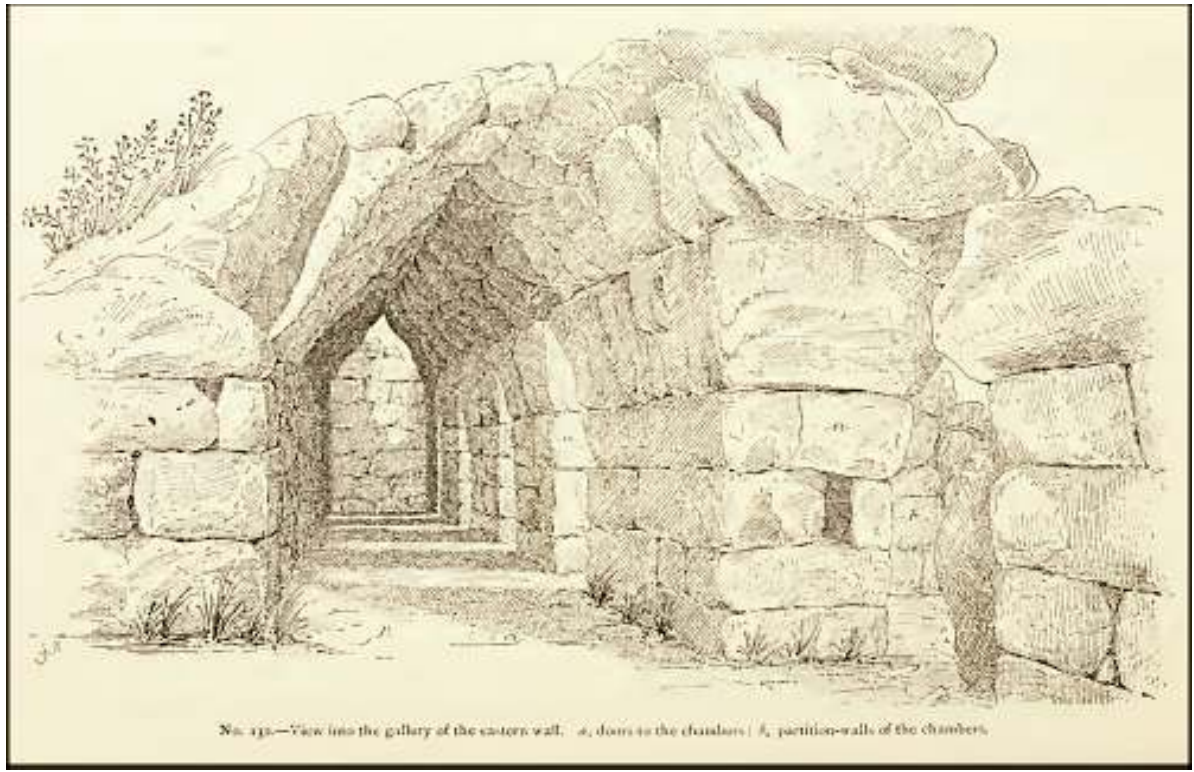

Fig.3: A tunnel in the cyclopean walls of Tiryns [13].

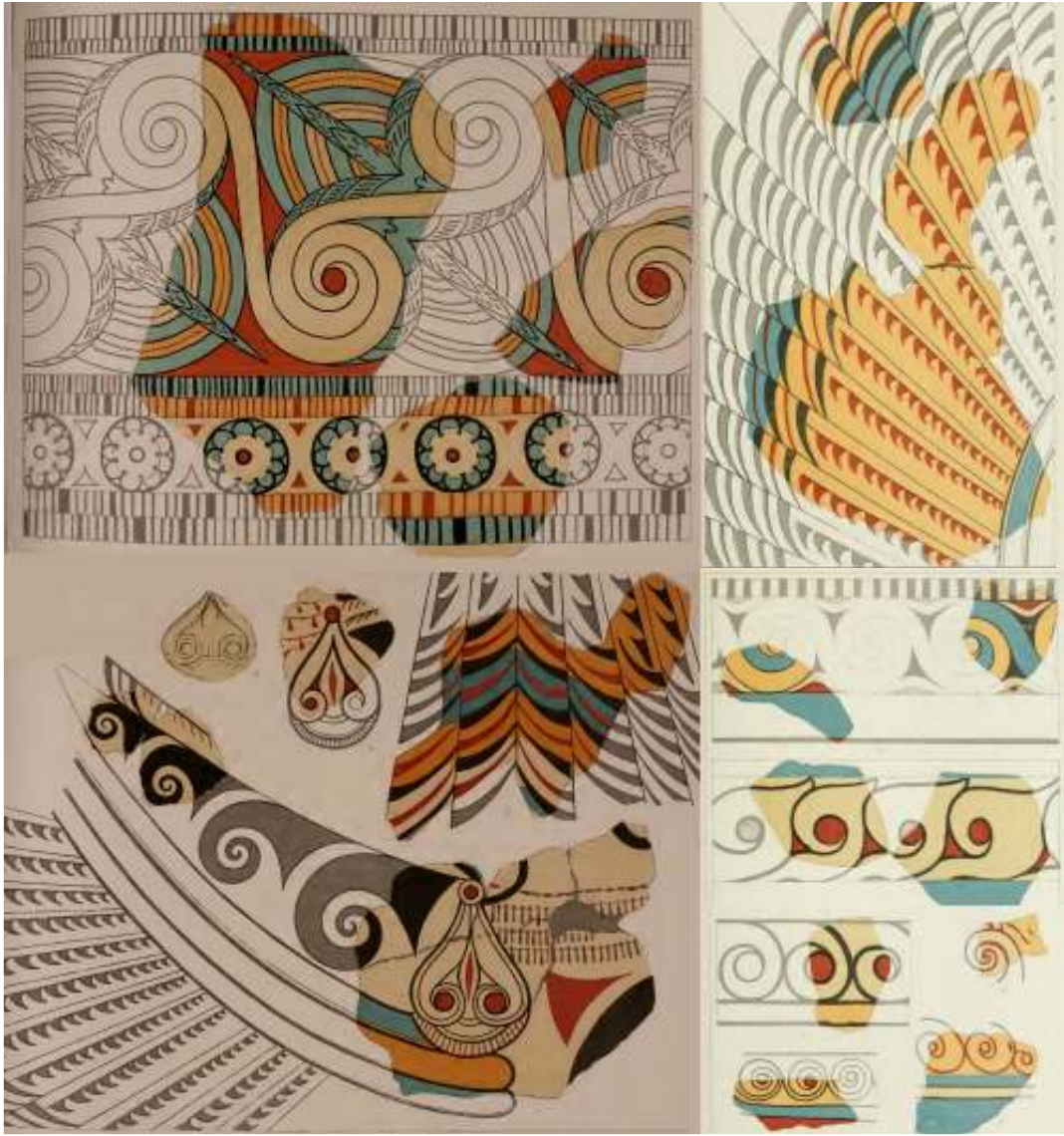

Fig.4: Decorations on the walls of the palace of Tiryns [13].